\documentstyle[aps,prb,tighten,eqsecnum]{revtex}
\begin{document}

\title{Faraday effect in composites}

\author{Marc Barth\'el\'emy}
\address{Commissariat \`a l'Energie Atomique, Service de Physique de la Mati\`ere
Condens\'ee\\
 BP 12, 91680
Bruy\`eres-Le-Ch\^atel, France}
\author{David J. Bergman}
\address{School of Physics and Astronomy, Raymond and Beverly Sackler
Faculty
of Exact Sciences\\Tel Aviv University, Tel Aviv 69978, Israel }

\date{\today}

\maketitle

\def\be{\begin{equation}}
\def\ee{\end{equation}}
\def\vr{\vec r}
\def\vk{\vec k}
\newcommand{\vrp}{\vec r'}
\newcommand{\gti}{\tilde g}
\newcommand{\ddk}{\frac{d^d k}{(2 \pi)^d}}

\vskip 1cm

\begin{abstract}
In this article, we study the Faraday effect in a binary composite
consisting of a dielectric matrix with metallic inclusions.
We first use the replica trick together with a variational
method in order to compute the effective permittivity tensor
(in the quasi-static limit) of this composite in a static magnetic
field. In order to find the scaling exponents near the percolation
threshold $p_c$, we use a
high contrast or low frequency expansion combined with scaling.
The results of the two methods are in agreement and predict that near
$p_c$ (and below, that is, in the dielectric region), the Faraday effect
is greatly enhanced.
\end{abstract}

\vskip 1cm\noindent\mbox{PACS: 72.15.Gd, 05.70.Jk, 72.80.Tm} \hfill
\vskip 1cm\noindent\mbox{Submitted for publication to: Phys. Rev. B}
\hfill


\section{Introduction}
\label{sec: intro}
We will concentrate on the problem of the effective
permittivity tensor of a binary mixture in a static
magnetic field. Since the Faraday effect is usually very weak
in dielectrics, we study a mixture made of a dielectric host
matrix (with a negligible Faraday effect) with metallic inclusions.
In such a system, the Hall effect in the inclusions is
expected to induce a measurable Faraday effect in the composite
dielectric even when this effect is totally negligible in the
pure dielectric host.
We will study the quasi-static limit, the wavelength of
the incoming wave being much larger than the typical
inhomogeneity length (which is for example the size of
the metallic grains). Therefore, the composite medium can
be seen as quasi-homogeneous, and the equivalent homogeneous
material is called the effective medium. Determining the
effective medium properties of disordered materials
(such as composites, suspensions) is a difficult problem
and one has often to resort to perturbative methods
(low field, low density or low contrast) which cannot
be applied for high magnetic field for instance.
We propose here two different approaches to this problem.
The first one is based on the use of the replica method together
with a variational principle. This treatment possesses the
advantage that it is non-perturbative and may thus be useful
for strong disorder or strong fields. This method was
successfully applied to different problems (the random
resistor network problem \cite{bo}, Hall effect in
composites \cite{bbo}). Moreover it has been shown that
it can give reasonable values for the permittivity tensor
if the system is not too close to the percolation threshold
$p_c$. However, the critical exponents are not reproduced
correctly (one usually
gets mean-field or effective-medium-approximation exponents).
In order to present an alternative discussion of this
problem, and obtain the correct exponents near
$p_c$, we introduce a high contrast expansion.
This is essentially an expansion in powers of the ratio of
resistivities or permittivities of the two components, which
can be made very small by making the frequency of the incoming wave
very small. This expansion can be used for weak magnetic fields
as well as for strong magnetic fields. In order to discuss the
critical properties near $p_c$, we apply some scaling ansatzes
to that expansion, which are based upon previous discussions
of d.c.\ magneto-transport \cite{BergmanStroud85,SarBergStrelPRB93}.

The Hall effect in percolating composites has been studied using a
number of different methods which enabled the critical behavior
to be determined quite reliably
\cite{Straley,StraleyLett,Duering,BergmanStroud85,bbo}.
However, until now the Faraday effect in metal-dielectric composites
was only discussed using a Clausius-Mossotti-type approximation
\cite{HuiStroudAPL87}, which is good for dilute systems, and
a Bruggeman-type self consistent effective medium approximation
(SEMA) \cite{XiaHuiStroudJAP90}, which exhibits a percolation
threshold but with incorrect values of the critical exponents.
In the present study we employ different
approaches (see above). Both of our approaches are not limited
to dilute systems, and one (the high contrast expansion together
with scaling ansatzes) is expected to lead to reliable results
for the critical behavior near $p_c$.

Let us first recall some facts about the Faraday effect. When an
isotropic material is subjected to a static and uniform magnetic
field {\bf B} directed along the $z$-axis, it can be described
by the following permittivity tensor:
\begin{equation}
\label{matrice}
\hat{\varepsilon}=\left(\begin{array}{ccc}
\varepsilon & i\tilde{\varepsilon} & 0\\
-i\tilde{\varepsilon} & \varepsilon & 0\\
0  & 0 & \varepsilon_{z}
\end{array}\right),\label{hat_eps}
\end{equation}
where $\tilde{\varepsilon}$ must depend upon {\bf B}
(see, e.g., Ref.\ \onlinecite{landau}).
In a homogeneous medium, the dispersion equation gives rise to two
solutions which are the left and right circularly polarized waves with two
different refractive indices ($\varepsilon$ and $\tilde{\varepsilon}$
are positive
and real and $\varepsilon >\tilde{\varepsilon}$, which ensures that
the wave is undamped)
\begin{equation}
\label{indices}
n_{\pm}=\sqrt{\varepsilon\pm\tilde{\varepsilon}}.
\end{equation}
If a linearly polarized wave of frequency $\omega$ propagates over
a distance $L$ through
this medium, the polarization plane will rotate (the
so-called Faraday effect) by an angle ($c$ is the light speed in vacuum)
\begin{equation}
\label{theta1}
\theta={\omega\over{c}} L(n_{+}-n_{-}).
\end{equation}
Usually $\tilde{\varepsilon}$ is very small compared to $\varepsilon$
(i.e., the Faraday effect is weak)
and the rotation angle is therefore approximately given by
\begin{equation}
\label{theta2}
\theta\simeq{\omega\over{c}}L{\tilde{\varepsilon}\over{\sqrt{\varepsilon}}}.
\end{equation}
We will also use the Faraday coefficient, which is defined by
${\cal F}\equiv\tilde{\varepsilon}/\sqrt{\varepsilon}$. We note
here that the Faraday effect is usually weak (${\cal F}$ ranges
from $10^{-6}$ in dielectrics to $10^{-2}$ for thin films of metallic
iron).

We will study the case where the material is a random binary
composite medium made of a dielectric host with metallic
inclusions, and where a static uniform magnetic field
{\bf B} is applied along the $z$-axis. We suppose that the medium
has a position dependent permittivity tensor
 $\hat{\varepsilon}({\bf r})$ which is an independent random variable
at each point ${\bf r}$ (uncorrelated at different positions),
distributed according to the following probability density
\begin{equation}
\label{pofeps}
p(\hat{\varepsilon})=p\delta(\hat{\varepsilon}
-\hat{\varepsilon}_{M})+(1-p)\delta(\hat{\varepsilon}
-\hat{\varepsilon}_{I}).
\end{equation}
Let us note that in real materials, the grains have finite sizes
and that in a finite frequency calculation one should take this
into account. However, in the quasi-static limit, the
grain sizes are irrelevant and this simplified characterization
of the disorder [Eq. (\ref{pofeps})] is justified.

Equivalently, the local resistivity tensor $\hat\rho({\bf r})$ (related
to the permittivity tensor $\hat{\varepsilon}$ by
$\hat\rho=\frac{4\pi}{i\omega\hat\varepsilon}$) is a step function
that is equal to $\hat\rho_M$ inside the metal and to
$\hat\rho_I$ inside the dielectric component. It will be represented,
with the help of the appropriate characteristic functions
$\theta_M({\bf r})$, $\theta_I({\bf r})$, as
\begin{eqnarray}
\label{caract}
\hat\rho({\bf r})&=&\hat\rho_M\theta_M({\bf r}) +
\hat\rho_I\theta_I({\bf r}),\label{hat_rho}\\
\theta_M({\bf r})&=&1-\theta_I({\bf r})=\left\{\begin{array}{ll}
1\;\;\;\mbox{for {\bf r} inside the metal,}\\
0\;\;\;\mbox{otherwise}.\end{array}\right.
\end{eqnarray}
In the dielectric component, the permittivity is taken to be
\begin{equation}
\label{permit1}
\hat{\varepsilon}_{I}=\varepsilon_{I}\hat{I}.
\end{equation}
where $\varepsilon_{I}$, the dielectric constant of the host,
is a real scalar quantity and is independent of
{\bf B}. The metallic
component is non-percolating, and is characterized by a
free-electron-like resistivity tensor in the presence of a magnetic
field ${\bf B}\parallel z$ or, equivalently, by its permittivity tensor
\begin{equation}
\hat{\varepsilon}_M=\varepsilon_M\left(\begin{array}{ccc}
{1\over{1+H^2}} & {H\over{1+H^2}} & 0\\
-{H\over{1+H^2}} & {1\over{1+H^2}} & 0\\
0  & 0 & 1
\end{array}\right),\;\;\;H\equiv\omega_c\tau\propto|{\bf B}|,
\label{perm_tensor}
\end{equation}
where $\omega_c$ is the cyclotron frequency, $\tau$ is the
conductivity relaxation time, and $\varepsilon_{M}=\frac{4\pi\sigma_M}
{i\omega}$ is purely imaginary and independent of {\bf B}
($\sigma_M$ is the conductivity of the metallic component). We
assume that this form continues to be valid even at finite frequencies.
This probably means that the entire subsequent discussion will
not be valid for optical frequencies in the visible range. But
it will be relevant for frequencies up to, and including,
the microwave regime.

These assumptions mean that the host exhibits no
intrinsic Faraday effect and the metallic component has no
intrinsic magneto-resistance, only a Hall effect. In terms of
resistivity, the metallic component is characterized by a
free-electron-like resistivity tensor (obtained by inverting
$\hat{\varepsilon}_M$)
\begin{equation}
\hat\rho_M=
\rho_M
\left(\begin{array}{ccc}
1 & H & 0\\
-H & 1 & 0\\
0 & 0 & 1
\end{array}\right),
\end{equation}
and the impedance of the dielectric host is an imaginary scalar
tensor
\begin{equation}
\hat\rho_I=\rho_I\hat I,\;\;\;\rho_I=\frac{4\pi}{i\omega\varepsilon_I}.
\end{equation}

We will assume that the quasi-static approximation can
be used (i.e., both wavelength and skin depth are large
compared to the sizes of metallic inclusions).

We define the bulk effective permittivity tensor $\hat{\varepsilon}_e$ of
the medium by the following relation
\begin{equation}
\label{epseff}
\langle\hat{\varepsilon}(r){\bf E}(r)\rangle=
\hat{\varepsilon}_{e}\langle{\bf E}(r)\rangle,
\end{equation}
where {\bf E} is the electric field and where the brackets denote
a quenched average over the probability distribution given by
(\ref{pofeps}) (or equivalently a
spatial average over the volume of the sample). The effective medium will
be homogeneous and isotropic
and we expect an effective permittivity tensor of the form
\begin{equation}
\label{matrixeff}
\hat{\varepsilon}_{e}=\left(\begin{array}{ccc}
\varepsilon_{e} & i\tilde{\varepsilon}_{e} & 0\\
-i\tilde{\varepsilon}_{e} & \varepsilon_{e} & 0\\
0  & 0 & \varepsilon_{ze}
\end{array}\right).
\end{equation}
We can also define the bulk effective complex resistivity tensor by
\begin{equation}
\label{rhoeff}
\hat\rho_e\cdot\langle{\bf J}\rangle\equiv
\langle\hat\rho{\bf J}\rangle,
\end{equation}
where {\bf J} is the local current density.

In order to evaluate the effective properties of the heterogeneous medium,
we can proceed in different ways.
The first one (which will be presented in Section \ref{sec: ra}) relies on
the observation that the effective permittivity tensor can be (exactly)
related
to the inverse of a random operator $\hat{M}$. The problem is thus reduced
to the calculation of $\langle\hat{M}^{-1}\rangle$, and we will use
replicas (and a variational approximation) to evaluate this quantity.
Let us note here that it is in principle equivalent to compute
$\hat{\varepsilon}_e$ using (\ref{epseff}) or
$\hat{\rho}_e$ using (\ref{rhoeff}), since the product of these tensors
is proportional to the unit tensor $\hat{I}$. However,
since we use approximations, the two procedures are not necessarily
equivalent. In fact, it has been shown \cite{hori,yone}
that for the variational approximation, computing $\hat{\varepsilon}_e$
or $\hat{\rho}_e$ leads to different results, and that it is more reliable
to calculate the permittivity (or the conductivity) than it is
to calculate the resistivity.

Since the variational method is non-perturbative, it possesses the
advantage
 that it is reliable in the whole range of concentration and for any
strength of the disorder. However, near
the percolation threshold $p_c$, this method leads to mean-field
exponents which are usually not accurate (especially
in three dimensions $d=3$). Moreover, it is difficult to control the
quality of the variational approximation.

In order to describe the Faraday effect near $p_c$, we
will use
a second approach, which relies on the fact that if $\omega$ is small
enough,
we have two small parameters $\rho_M/\rho_I$ and $H\rho_M/\rho_I$.
The local electric field ${\bf E}({\bf r})$
and current density ${\bf J}({\bf r})$ can be found by defining
a vector potential ${\bf A}({\bf r})$ such that
\begin{equation}
{\bf J}({\bf r})=\nabla\times{\bf A}({\bf r}),
\end{equation}
and then solving the equation
\begin{equation}
\nabla\times{\bf E}=0
\end{equation}
using the constitutive relation ${\bf E}({\bf r})=
\hat{\rho}({\bf r}){\bf J}({\bf r})$. It is of course
out of the question to solve this equation exactly,
therefore we will  expand {\bf A}
in powers of the contrast between the two components, i.e.,
the resistivity ratio $\rho_M/\rho_I$.
This approach is valid only in the quasi-static
regime. Besides confirming results obtained by means of
the replicas, when combined with scaling it allows us to predict the
behavior of
the medium near the percolation threshold.

The rest of this paper is organized as follows. In Section \ref{sec:
ra} we apply the replica method to the Faraday effect in random composites. Results are obtained for the scaling behaviors near the percolation threshold. In Section \ref{sec: st} we present the high contrast or low
frequency expansion. Section \ref{scaling} presents a scaling theory based upon the
approaches described in the two previous sections. Section
\ref{conclusion}
summarizes the main conclusions from this work.


\section{Replica approach}
\label{sec: ra}

In this section we present the replica approach. We compute here the
effective permittivity tensor of a binary mixture, where the tensor
$\hat{\varepsilon}$ is a random variable equal to $\hat{\varepsilon}_1$
with probability $p$ and to $\hat{\varepsilon}_2$ with probability $q=1-p$
(each component having a permittivity of the form given in equation
(\ref{hat_eps})). The calculation is essentially the same as for the
Hall effect \cite{bbo} and we give the main steps of the derivation in
Appendix A. We obtain the following equations: The first one gives the
longitudinal effective permittivity
\begin{equation}
\label{epsz}
\varepsilon_{ze}=\int_0^\infty{\rm d}ue^{-u}
\frac
{\langle\varepsilon_{z}e^{-u\varepsilon_{z}/3
\varepsilon_{ze}}\rangle}
{\langle e^{-u\varepsilon_{z}/3\varepsilon_{ze}}\rangle},
\end{equation}
where the brackets still denote an average over (\ref{pofeps}),
and we also get two coupled equations for the transverse permittivities
$\varepsilon_e$, $\tilde{\varepsilon}_e$:
\begin{equation}
\label{epseff1}
1=-{3\over{2}}\int_0^\infty{\rm d}ue^{-u}
\{\ln\langle e^{-uX/3\varepsilon_e}\rangle
+\ln\langle e^{-uY/3\varepsilon_e}\rangle \},
\end{equation}
\begin{equation}
\label{epstilde}
\tilde{\varepsilon}_e=
{3\over{2}}\varepsilon_e
\int_0^\infty{\rm d}ue^{-u}
\{\ln\langle e^{-uX/3\varepsilon_e}\rangle
-\ln\langle e^{-uY/3\varepsilon_e}\rangle \},
\end{equation}
where $X\equiv\varepsilon-\tilde{\varepsilon}$ and
$Y\equiv\varepsilon+\tilde{\varepsilon}$.
Eq.\ (\ref{epseff1}) determines
$\varepsilon_e$ self-consistently while
(\ref{epstilde}) determines
$\tilde{\varepsilon}_e$ after $\varepsilon_e$ is known.

Eq. (\ref{epsz}) is the same as that obtained in Refs.\
\onlinecite{bo}
and \onlinecite{hori} for the bulk effective conductivity of
a binary mixture of zero-field conductivities $\sigma_{z1}$
with concentration $p$ and
$\sigma_{z2}$ with concentration $1-p$.
That equation was studied thoroughly in Ref.\ \onlinecite{yone};
it displays
a percolation threshold at $p_c=1-e^{-1/3}\simeq 0.28$.

The Faraday effect is contained in (\ref{epseff1}) and (\ref{epstilde}).
We first note that the percolation threshold is independant of the
magnetic field. Let us note here that the percolation threshold, which
is a geometrical quantity, is still meaningful here since we are working
in the quasi-static
limit.

By expanding the logarithms (these expansions are valid for $p<1/2$),
we obtain from (\ref{epseff1})
\begin{equation}
\label{alphacont}
1=-3{\rm ln}q+\frac{\varepsilon_2}{\varepsilon_e}
-\frac{3}{2}\sum_{n\ge 1}\frac{\lambda^n}{n}(-)^{n+1}
\left(\frac{1}{1+n\frac{(X_1-X_2)}{3\varepsilon_e}}+
\frac{1}{1+n\frac{(Y_1-Y_2)}{3\varepsilon_e}}\right),
\end{equation}
and from (\ref{epstilde})
\begin{equation}
\label{betacont}
\tilde{\varepsilon}_e=\tilde{\varepsilon}_2
+(\tilde{\varepsilon}_1-\tilde{\varepsilon}_2)
\sum_{n\ge 1}\lambda^n(-)^{n+1}
\frac{1}{
[1+{n\over{3\varepsilon_e}}(X_1-X_2)]
[1+{n\over{3\varepsilon_e}}(Y_1-Y_2)] },
\end{equation}
where $\lambda=p/q$, with $p$ the fraction of component $1$ and $q=1-p$
the fraction of component $2$.
By some simple algebraic manipulations, these equations
can be rewritten as
\begin{equation}
\label{alpha}
0=\frac{1}{3}+{\rm ln}q-\frac{\alpha_2}{3\alpha_e}+
\sum_{n\ge 1}\frac{\lambda^n}{n}(-)^{n+1}
\frac{ 1+\frac{n(\alpha_1-\alpha_2)}{3\alpha_e} }
{ (1+\frac{n(\alpha_1-\alpha_2)}{3\alpha_e})^2+
(\frac{n(\beta_1-\beta_2)}{3\alpha_e})^2 }
\end{equation}
and
\begin{equation}
\label{beta}
\beta_e=\beta_2+(\beta_1-\beta_2)
\sum_{n\ge 1}\frac{\lambda^n}{n}(-)^{n+1}
\frac{1}
{ (1+\frac{n(\alpha_1-\alpha_2)}{3\alpha_e})^2+
(\frac{n(\beta_1-\beta_2)}{3\alpha_e})^2 },
\end{equation}
where $\alpha_i$ and $\beta_i$
describe, respectively, the diagonal and the off-diagonal transverse
elements of the
conductivity tensor of component $i$ [$\alpha_i=\sigma_i/(1+H_i^2)$,
$\beta_i=\sigma_i H_i/(1+H_i^2)$],
and where $\sigma_i$ is the zero-field conductivity of
component $i$ [$H_i$ is the Hall-to-Ohmic resistivity ratio
 in component $i$: it is proportional to the magnetic field strength
$|{\bf B}|$---see (\ref{perm_tensor})]. The
quantities $\alpha_e$ and $\beta_e$ are the effective
coefficients of the composite. Eqs.\ (\ref{alpha}) and (\ref{beta}) are
identical to the equations obtained in the case of magneto-transport
\cite{bbo}. We have thus shown that the results obtained in that case
can be continued analytically to complex values of $\alpha$ and $\beta$
(we can go from the Hall effect to the Faraday effect by changing $\alpha$
to $\varepsilon$ and $\beta$ into $i\tilde{\varepsilon}$).

We now study these equations in the case of a metal-dielectric
mixture (the metal will be component $1$ and the dielectric component
$2$). We assume here that the Hall effect in the dielectric
is very weak so that $\alpha_2/\alpha_1\ll 1$ or $H_2\simeq 0$,
and we use $H$ instead of $H_1$. We also assume
that $\alpha_e\ll\alpha_1$.

We will first study the weak field regime $H\ll 1$, in this case
$\beta_i\ll\alpha_i$. Near the percolation threshold $p_c=1-e^{-1/3}$ of
component $1$ ($\Delta p=p-p_c\ll 1$), we find that $\alpha_e$
has the following scaling behavior $\alpha_e\simeq\alpha_1|\Delta
p|\phi(\frac{\alpha_2/\alpha_1}{\Delta p^2})$,
where the scaling function $\phi(z)$ satisfies
\begin{equation}
0\simeq -\frac{\Delta p}{q_c}-\frac{\alpha_2}{3\alpha_1|\Delta p|\phi}
+|\Delta p|\phi A,
\end{equation}
where $A=3\sum_{n\ge 1}\frac{\lambda_c^n}{n^2}(-)^{n+1}$ (with
$\lambda_c=p_c/1-p_c$).
We thus obtain the following equation for $\phi$
\begin{equation}
A\phi^2-\frac{\epsilon\phi}{q_c}-\frac{z}{3}=0,
\end{equation}
where $\epsilon=+1$ if $p>p_c$ and $\epsilon=-1$ for $p<p_c$, and where
$z=\frac{\alpha_2}{\alpha_1\Delta p^2}$. The solution of this equation is
$\phi=\frac{1}{2A}(\frac{\epsilon}{q_c}+\sqrt{\frac{1}{q_c^2}+\frac{4Az}{3}})$,
which for small $z$ becomes (up to a constant factor $q_c/3$)
\begin{eqnarray}
\label{F}
\phi(z)&\propto&
\left\{\begin{array}{lll}
z-\frac{A}{3}q_c^2z^2; & p<p_c & \\
Const; & p>p_c. & \end{array}\right.
\end{eqnarray}
We can now easily obtain the behavior of $\beta_e$ from
(\ref{beta}) (for $p<p_c$)
\begin{equation}
\label{betalowH}
\beta_e\propto\frac{\beta_1}{\alpha_1^2}\alpha_e^2
\propto\beta_1(\frac{\alpha_2/\alpha_1}{|\Delta p|})^2,
\end{equation}
which is proportional to $H$.

We now consider the regime $H\gg 1$ for $p$ below $p_c$ (which
is the interesting regime for the following). In this regime
$\beta_i\gg\alpha_i$, and (\ref{alpha}) then reads
\be
0\simeq -\frac{\Delta
p}{q_c}-\frac{\alpha_2}{3\alpha_e}+\frac{\alpha_1\alpha_e}{\beta_1^2}A,
\end{equation}
from which we can deduce that the scaling behavior of $\alpha_e$ is of
the form
$\alpha_e\simeq\frac{\beta_1^2}{\alpha_1}|\Delta
p|\tilde{\phi}(\frac{\alpha_2\alpha_1}{\beta_1^2\Delta p^2})$, where the
scaling function $\tilde{\phi}(z)$ has a behavior like that
of $\phi(z)$, up to the same constant factor $q_c/3$ [see (\ref{F})]:
\begin{eqnarray}
\label{Ftilde}
\tilde{\phi}(z)&\propto&
\left\{\begin{array}{lll}
z-\frac{A}{3}q_c^2z^2; & p<p_c & \\
Const; & p>p_c. & \end{array}\right.
\end{eqnarray}
The behavior of $\beta_e$ is {\it a priori} different,
since now $\beta_1\propto\frac{1}{H}\gg\alpha_1\propto\frac{1}{H^2}$.
The equation for $\beta_e$ then becomes
\begin{equation}
\beta_e\simeq\beta_1\sum_{n\ge 1}\frac{\lambda_c^n}{n^3}(-)^{n+1}
\frac{1}{(\beta_1/3\alpha_e)^2}
\end{equation}
which for $p<p_c$ leads to the following behavior
\begin{equation}
\label{betahighH}
\beta_e\propto\frac{\alpha_2^2}{\Delta p^2}\frac{1}{\beta_1},
\end{equation}
which is again proportional to $H$.

We will discuss the physical consequences relevant to the Faraday
effect in Section \ref{scaling}.


\section{Low frequency or high contrast expansion}

\label{sec: st}

In this section we derive an expansion for $\hat\varepsilon_e$
in powers of the complex resistivity ratio
$\rho_M/\rho_I$. Throughout this section we will assume that
$\omega$ is small enough so that both
$\rho_I\gg\rho_M$ and $\rho_I\gg H\rho_M$, and that the
quasi-static limit is valid.

Under these assumptions, the local electric field ${\bf E}({\bf r})$
and current density ${\bf J}({\bf r})$ can be found by defining
a vector potential ${\bf A}({\bf r})$ such that
\begin{equation}
{\bf J}({\bf r})=\nabla\times{\bf A}({\bf r}),\label{A_def}
\end{equation}
and solving the equation
\begin{equation}
\label{A_diff_eq}
0=\nabla\times{\bf E}=\nabla\times\{\hat\rho({\bf r})\cdot
[\nabla\times{\bf A}({\bf r})]\}
\end{equation}
along with appropriate boundary conditions on
${\bf n}\times{\bf A}({\bf r})$ at the system surface
({\bf n} is the unit normal vector to that surface).

We recall that the local resistivity tensor $\hat\rho({\bf r})$ is a
step function
equal to $\hat{\rho}_M$ inside the metal and to $\hat{\rho}_I$
inside the dielectric component, and that it can be represented with the
help of the appropriate characteristic functions as in
(\ref{caract}).

In connection with (\ref{A_diff_eq}) it is useful to define
a Green tensor $\hat G^{(\rho)}({\bf r},{\bf r'})$ by the following
equations
\begin{eqnarray}
\{\nabla\times[\hat\rho\cdot(\nabla\times
G^{(\rho)}_{\cdot\beta})]\}_\alpha
-k^2G^{(\rho)}_{\alpha\beta}&=&\delta_{\alpha\beta}\delta^3({\bf
r}-{\bf r'})
\;\;\;\mbox{for any}\;\; {\bf r}, {\bf r'},\label{G_eq}\\
{\bf n}\times G^{(\rho)}_{\cdot\beta}&=&0\;\;\;\mbox{for {\bf r}
at the system surface}.
\label{G_bound_cond}
\end{eqnarray}
This tensor can be used to solve (\ref{A_diff_eq}):
\begin{equation}
{\bf A}({\bf r})={\bf A}^{(0)}({\bf r}) -
\int{\rm d}{\bf r'}\lim_{k\rightarrow 0}[\nabla'\times\hat G^{(\rho)}({\bf
r},{\bf r'})]
\cdot\hat\rho({\bf r'})\cdot[\nabla'\times{\bf A}^{(0)}({\bf r'})],
\label{A_solution}
\end{equation}
where ${\bf A}^{(0)}({\bf r})$ is a vector
field that satisfies the same boundary conditions as
${\bf A}({\bf r})$, but is otherwise arbitrary. Note that we
need to use the limit $k\rightarrow 0$ of
$\nabla'\times\hat G^{(\rho)}({\bf r},{\bf r'})$ here:
We could not take that limit in (\ref{G_eq}), because then
the equations for $\hat G^{(\rho)}$  would have no solution
(see Appendix B for a discussion of this point).

Since we intend to expand {\bf A} in powers of $\rho_M/\rho_I$,
we define
\begin{equation}
\hat G^{(I)}\equiv\lim_{\rho_M\rightarrow 0}\hat G^{(\rho)}.
\end{equation}
It is then possible to transform (\ref{G_eq}), (\ref{G_bound_cond})
into an integro-differential equation that relates $\hat G^{(I)}$
and $\hat G^{(\rho)}$:
\begin{equation}
G^{(\rho)}_{\alpha\beta}({\bf r}, {\bf r'})=
G^{(I)}_{\alpha\beta}({\bf r}, {\bf r'})-
\int{\rm d}r^{\prime\prime}[\nabla^{\prime\prime}\times
G^{(I)}_{\alpha\cdot}
({\bf r},{\bf r^{\prime\prime}})]
\cdot\hat\rho_M\theta_M({\bf r^{\prime\prime}})\cdot[
\nabla^{\prime\prime}\times G^{(\rho)}_{\cdot\beta}({\bf
r^{\prime\prime}},
{\bf r'})].\label{G_integral_eq}
\end{equation}
Iteration of this equation leads, in the usual way, to an expansion
of $\hat G^{(\rho)}$ in powers of $\hat\rho_M$ around $\hat G^{(I)}$.
We note that, although
$G^{(\rho)}_{\alpha\beta}({\bf r},{\bf r'})$ is not a symmetric kernel
(because $\hat\rho({\bf r})$ is a non-symmetric tensor),
$G^{(I)}_{\alpha\beta}({\bf r},{\bf r'})$ is symmetric because
$\hat\rho_I$ is symmetric (in fact, $\hat\rho_I$ is a scalar
tensor) [see Appendix B for a discussion of this point]
\begin{equation}
G^{(I)}_{\alpha\beta}({\bf r},{\bf r'})=
G^{(I)}_{\beta\alpha}({\bf r'},{\bf r}).\label{G_I_symm}
\end{equation}

A possible choice of ${\bf A}^{(0)}$ in (\ref{A_solution})
is $\lim_{\rho_M\rightarrow 0} {\bf A}$---henceforth we adopt that choice.
If we then take
the limit $\rho_M\rightarrow 0$ also in the functions
${\bf A}({\bf r})$, $\hat G^{(\rho)}({\bf r},{\bf r'})$, we
conclude that
\begin{equation}
\int {\rm d}{\bf r}'\lim_{k\rightarrow 0}[\nabla'\times\hat G^{(I)}
({\bf r},{\bf r'})]
\cdot\hat\rho_I\theta_I({\bf r'})\cdot[\nabla'\times{\bf A}^{(0)}({\bf
r'})]
=0.
\end{equation}
Using the above mentioned power series expansion for $\hat G^{(\rho)}$,
this result can be extended to hold also when $\hat G^{(I)}$ is
replaced by $\hat G^{(\rho)}$:
\begin{equation}
\int{\rm d}{\bf r}'\lim_{k\rightarrow 0}[\nabla'\times\hat G^{(\rho)}({\bf
r},{\bf r'})]
\cdot\hat\rho_I\theta_I({\bf r'})\cdot[\nabla'\times{\bf A}^{(0)}({\bf
r'})]
=0.
\end{equation}
Using this result together with (\ref{G_integral_eq}),
(\ref{A_solution}) can be transformed into an expansion for
${\bf A}({\bf r})$ in powers of $\rho_M$:
\begin{eqnarray}
{\bf A}({\bf r})&=&{\bf A}^{(0)}({\bf r}) -
\int{\rm d}{\bf r'}\lim_{k\rightarrow 0}[\nabla'\times\hat G^{(\rho)}({\bf
r},{\bf r'})]
\cdot\hat\rho_M\theta_M({\bf r'})\cdot[\nabla'\times{\bf A}^{(0)}({\bf
r'})]
\nonumber\\&=&{\bf A}^{(0)}({\bf r})-
\int{\rm d}{\bf r'}\lim_{k\rightarrow 0}[\nabla'\times\hat G^{(I)}({\bf
r},{\bf r'})]
\cdot\hat\rho_M\theta_M({\bf r'})\cdot[\nabla'\times{\bf A}^{(0)}({\bf
r'})]
\nonumber\\&&\;\;\;+\,O(\rho_M^2).\label{A_solution_1}
\end{eqnarray}
We also note that ${\bf A}^{(0)}({\bf r})$ can be obtained by
an expression that is the analogue of (\ref{A_solution}), namely
\begin{equation}
{\bf A}^{(0)}({\bf r})={\bf A}^{(00)}({\bf r}) -
\int{\rm d}{\bf r'}\lim_{k\rightarrow 0}[\nabla'\times\hat G^{(I)}({\bf
r},{\bf r'})]
\cdot\hat\rho_I\theta_I({\bf r'})\cdot[\nabla'\times{\bf A}^{(00)}({\bf
r'})].
\label{A_0_solution}
\end{equation}
This was obtained by replacing ${\bf A}^{(0)}({\bf r})$ by
${\bf A}^{(00)}({\bf r})$ in (\ref{A_solution}), and then taking
the limit $\rho_M\rightarrow 0$ in that equation. Eq. (\ref{A_0_solution})
is especially useful if we assume the following boundary
condition for ${\bf A}({\bf r})$ and ${\bf A}^{(0)}({\bf r})$
\begin{equation}
{\bf A}({\bf r})={\bf A}^{(0)}({\bf r})=\frac{1}{3}({\bf e}\times{\bf r})
\;\;\;\mbox{at the system surface,}\label{e_bound_cond}
\end{equation}
and choose
\begin{equation}
{\bf A}^{(00)}({\bf r})=\frac{1}{3}({\bf e}\times{\bf r})\;\;\;
\mbox{everywhere,}
\end{equation}
where {\bf e} is some unit vector. This last choice corresponds to
a uniform
current density
$$
\nabla\times\left[\frac{1}{3}({\bf e}\times{\bf r})\right]={\bf e},
$$
which is equal to the {\em volume averaged} current density for
both ${\bf A}^{({\bf e})}({\bf r})$ and ${\bf A}^{(0{\bf e})}({\bf r})$,
which satisfy (\ref{e_bound_cond}),
\begin{equation}
\langle\nabla\times{\bf A}^{({\bf e})}\rangle=
\langle\nabla\times{\bf A}^{(0{\bf e})}\rangle={\bf e}.\label{A_e_average}
\end{equation}
We thus find that ${\bf A}^{(0{\bf e})}$ is given by
\begin{equation}
{\bf A}^{(0{\bf e})}({\bf r})=\frac{1}{3}({\bf e}\times{\bf r}) -
\int{\rm d}{\bf r}'\lim_{k\rightarrow 0}[\nabla'\times\hat G^{(I)}({\bf
r},{\bf r'})]
\cdot\hat\rho_I\theta_I({\bf r'})\cdot{\bf e}.\label{A_e_0}
\end{equation}

We recall that the bulk effective complex resistivity tensor of the system
is defined by [see (\ref{rhoeff}) and \ref{A_def})]
\begin{equation}
\hat\rho_e\cdot\langle\nabla\times{\bf A}\rangle\equiv
\langle\hat\rho\cdot(\nabla\times{\bf A})\rangle,
\end{equation}
for any {\bf A} of the form (\ref{A_solution}). Using
(\ref{hat_rho}) and (\ref{A_solution_1}), we can expand
an arbitrary element of the tensor $\hat\rho_e$ in powers of
$\rho_M$ ({\bf f}, {\bf e} are arbitrary unit vectors)
\begin{eqnarray}
({\bf f}\cdot\hat\rho_e\cdot{\bf e})&=&
{\bf f}\cdot\langle\hat\rho\cdot(\nabla\times{\bf A}^{({\bf e})})\rangle
\nonumber\\&=&
{\bf f}\cdot\hat\rho_I\cdot\langle\theta_I(\nabla\times{\bf A}^{(0{\bf
e})})\rangle
+{\bf f}\cdot\hat\rho_M\cdot\langle\theta_M(\nabla\times{\bf A}^{(0{\bf
e})})\rangle
\nonumber\\&&
-\,{\bf f}\cdot\frac{1}{V}\int{\rm d}{\bf r}\,\hat\rho_I\theta_I({\bf
r})\cdot
\nabla\times\nonumber\\&&
\int{\rm d}{\bf r}'\lim_{k\rightarrow 0}[\nabla'\times\hat G^{(I)}({\bf
r},{\bf r'})]
\cdot\hat\rho_M\theta_M({\bf r'})\cdot
[\nabla'\times{\bf A}^{(0{\bf e})}({\bf r'})]\nonumber\\&&\;\;\;
 +\, O(\rho_M^2).\label{rho_e_1}
\end{eqnarray}
(note that $\langle\theta_I(\nabla\times{\bf A}^{(0{\bf
e})})\rangle$ is just the spatial average of $\nabla\times{\bf A}^{(0{\bf e})}$
over the subvolume of the dielectric component, while
$\langle\theta_M(\nabla\times{\bf A}^{(0{\bf
e})})\rangle$ is the average of the same quantity over the metallic subvolume).
The integration over ${\bf r}$ can be performed using (\ref{G_I_symm})
and (\ref{A_e_0}), leading to the following result for the double integral
of (\ref{rho_e_1})
\begin{equation}
\int{d}{\bf r}'[\nabla'\times{\bf A}^{(0{\bf f})}({\bf r'})-{\bf f}]\cdot
\hat\rho_M\theta_M({\bf r'})\cdot
[\nabla'\times{\bf A}^{(0{\bf e})}({\bf r'})].
\end{equation}
Part of this cancels the second term of (\ref{rho_e_1}), and we
finally get
\begin{eqnarray}
{\bf f}\cdot\hat\rho_e(\hat\rho_I,\hat\rho_M)\cdot{\bf e}&=&
{\bf f}\cdot\hat\rho_I\cdot\langle\theta_I(\nabla\times{\bf A}^{(0{\bf
e})})\rangle
\nonumber\\
&&+\,\langle\theta_M(\nabla\times{\bf A}^{(0{\bf f})})\cdot\hat\rho_M\cdot
(\nabla\times{\bf A}^{(0{\bf e})})\rangle
\, +\, O(\rho_M^2).\label{rho_e_I_M}
\end{eqnarray}

If the microstructure is isotropic, then since $\hat\rho_I$ is
a scalar tensor, $\hat\rho_e(\hat\rho_I,0)$ [the first term on
the r.h.s.\ of (\ref{rho_e_I_M})] is also a scalar
tensor, and it is clearly independent of $\hat\rho_M$ and
hence of $H$.
The vector potentials ${\bf A}^{(0{\bf e})}$, ${\bf A}^{(0{\bf f})}$,
which appear in the second term on the r.h.s.\ of (\ref{rho_e_I_M}), satisfy
different boundary conditions {\em at the system surface}\, [see
(\ref{A_e_average})]. Inside
the metallic subvolume, those
potentials can also be viewed as resulting from boundary conditions
on ${\bf n}\times{\bf A}^{(0)}$ {\em at the interface
between the two components}\,. Those latter boundary values
are entirely determined by the microstructure when we impose
the requirement that the electric potential must be constant
over every connected subvolume of the metallic component, but the
precise local
values of ${\bf A}^{(0)}({\bf r})$ inside those subvolumes
also depend upon the Hall-to-Ohmic resistivity ratio of
the metal $H$. Nevertheless, we now argue that even the second
volume average
which appears in (\ref{rho_e_I_M}) is independent of $H$ in
the two limits $H\ll 1$ and $H\gg 1$. The only $H$ dependence in
those limits arises from the explicit $\hat\rho_M$ factor in
that term.

In order to prove this, we note that if the resistivity ratio
$\rho_M/\rho_I$ is small enough, then the current distribution
inside the metallic subvolumes, though different for $H\ll 1$ and for
$H\gg 1$,
will be saturated in both limits: In the weak field limit this
is obvious, while in the strong field limit this holds in
a percolating system whenever the magnetic field dependent correlation
length $\xi_H$, which diverges as $H\rightarrow\infty$,
is greater than the percolation correlation length\cite{SarBergStrelPRB93}
$\xi_p$.

Recalling that ${\bf B}\parallel z$, and assuming that the
microstructure is either isotropic or cubic, we now get that the
diagonal elements of $\hat\rho_e$ are given by
\begin{equation}
\rho^{(e)}_{\alpha\alpha}(\hat\rho_I,\hat\rho_M)=\rho_I
\langle\theta_I(\nabla\times{\bf A}^{(0\alpha)})_\alpha\rangle +
\rho_M\langle\theta_M(\nabla\times{\bf A}^{(0\alpha)})^2\rangle
+O(\rho_M^2),
\end{equation}
while the nonzero off-diagonal elements are
\begin{equation}
\rho^{(e)}_{xy}(\hat\rho_I,\hat\rho_M)=
-\rho^{(e)}_{yx}(\hat\rho_I,\hat\rho_M)=
H\rho_M\langle\theta_M[(\nabla\times{\bf A}^{(0x)})\times
(\nabla\times{\bf A}^{(0y)})]_z\rangle +O(\rho_M^2).
\end{equation}
Recalling also that
$$
\hat\rho_e=\frac{4\pi}{i\omega\hat\varepsilon_e},
$$
we finally get the following results for $\hat\varepsilon_e$
($\sigma_M\equiv 1/\rho_M$):
\begin{eqnarray}
\label{epstilda-st}
\hat\varepsilon_e(\varepsilon_I,\hat\rho_M)&\simeq&
\frac{\varepsilon_I}{\langle\theta_I(\nabla\times{\bf
A}^{(0x)})_x\rangle}\hat I
-\frac{i\omega\varepsilon_I^2}{4\pi\sigma_M}\frac{\langle\theta_M(\nabla
\times{\bf A}^{(0x)})^2\rangle}
{\langle\theta_I(\nabla\times{\bf A}^{(0x)})\rangle^2}\left(
\begin{array}{ccc}
1  &  0  &  0\\
0  &  1  &  0\\
 0  &  0  &  0\end{array}\right)\nonumber\\
&&-\,\frac{i\omega\varepsilon_I^2}{4\pi\sigma_M}\frac{\langle\theta_M(\nabla
\times{\bf A}^{(0z)})^2\rangle}
{\langle\theta_I(\nabla\times{\bf A}^{(0z)})\rangle^2}\left(
\begin{array}{ccc}
0  &  0  &  0\\
0  &  0  &  0\\
0  &  0  &  1\end{array}\right)\nonumber\\
&&-\,\frac{i\omega\varepsilon_I^2H}{4\pi\sigma_M}\,\frac{\langle\theta_M
[(\nabla\times{\bf A}^{(0x)})\times(\nabla\times{\bf A}^{(0y)})]_z\rangle}
{\langle\theta_I(\nabla\times{\bf A}^{(0x)})\rangle^2}
\left(\begin{array}{ccc}
0  & 1 & 0\\
-1 & 0 & 0\\
0  & 0 & 0
\end{array}\right).\nonumber\\&&\label{hat_epsilon_e}
\end{eqnarray}
This expression is the main result of this part and we will discuss
it in the next section.


\section{Scaling theory}
\label{scaling}

In this part, we will discuss physical consequences of the
results obtained by the different approaches, and we will use a
scaling theory in order to discuss the behavior
of the permittivity below the percolation threshold by comparing with
scaling
theories developed earlier for d.c. magneto-transport in a
percolating system
\cite{BergmanStroud85,SarBergStrelPRB93,BergStroudSSP}.

As we noted earlier, Eq.\ (\ref{epsz}) is the same as the result
obtained in
Refs.\ \onlinecite{bo}
and \onlinecite{hori}: It is an equation for the bulk
effective conductivity of a binary mixture which was
studied thoroughly in Ref.\ \onlinecite{yone}; in particular, it displays
a percolation threshold at $p_c=1-e^{-1/3}\simeq 0.28$.
The Faraday effect is contained in Eqs.\ (\ref{epseff1})
and (\ref{epstilde}) or,
equivalently, in (\ref{alpha}) and (\ref{beta}).
We first note that the percolation threshold for that effect
is independent of the magnetic field and, as expected, is the same as
the threshold found for the conductivity.

We can easily check those results in two limiting cases.
First, for zero magnetic field it is easy to see that one recovers
Hori's equation for the effective permittivity of a binary
mixture \cite{hori}. Then, for low concentration ($p\ll 1$) we find that
$\varepsilon_e\simeq\varepsilon_I$ and $\tilde{\varepsilon}_e\simeq
p{4\pi\sigma_M\over{\omega}}H/(1+H^2)$, and the rotation angle therefore
satisfies $\theta\propto 1/H$ for high fields and $\theta\propto
H$ for low fields. This agrees to order $O(p)$
with the low density expansion results for spherical inclusions
as obtained, for example, from the Clausius-Mossotti-type
approximation of Ref.\ \onlinecite{HuiStroudAPL87}.
We note that here $\theta$ is independent
of the frequency and is always very small.

We now apply our discussion from Section \ref{sec: ra}
to the case of a non-dilute metal-dielectric mixture:
For the metal,
$\alpha_1={\sigma_M\over(1+H^2)}$ and $\beta_1=\alpha_1 H$, and for
the dielectric (which is assumed to have a negligible Faraday effect)
$\alpha_2=i\omega\varepsilon_I/4\pi$ and $\beta_2\simeq 0$. We will
concentrate on the critical region near the percolation
threshold $p_c$, where $\Delta p\equiv
p-p_c$ is small ($p$ is the metal volume fraction).
We will now consider $\varepsilon_e$ and
$\tilde{\varepsilon}_e$ separately for the weak field regime ($H\ll 1$)
and the strong field regime ($H\gg 1$).

At low fields, we use the scaling result for
$\alpha_e$, and thus find that the effective permittivity
($\varepsilon_e=\frac{4\pi\alpha_e}{i\omega}$) is given by
\begin{equation}
\label{epselowH}
\varepsilon_e\simeq\frac{q_c}{3}\left(\frac{\varepsilon_I}{|\Delta
p|}-i\frac{Aq_c^2}{12\pi}
\frac{\omega\varepsilon_I^2}{\sigma_M|\Delta p|^3}\right)
\end{equation}
for the regime where $\frac{\omega\varepsilon_I}{\sigma_M\Delta
p^2}\ll 1$
and $p<p_c$.
The real part of $\varepsilon_e$ thus  diverges like $1/|\Delta p|$
(exponent $s$ equal to its SEMA value 1), and the imaginary part
is proportional to $\frac{\omega\varepsilon_I^2}{\sigma_M|\Delta p|^3}$.
>From (\ref{betalowH}), we obtain the following result for
$\tilde{\varepsilon}_e$
($\tilde{\varepsilon}_e=\frac{4\pi\beta_e}{\omega}$)
\be
\label{epstildelowH}
\tilde{\varepsilon}_e\propto -\frac{\omega\varepsilon_I^2}{4\pi\sigma_M
|\Delta p|^2}H.
\ee
In this regime, the Faraday coefficient will be
\be
\label{faradcoefflowH}
{\cal F}\propto -\frac{\omega\varepsilon_I^{3/2}H}{4\pi\sigma_M|\Delta
p|^{3/2}}.
\ee

In the strong field regime, $\beta_1\simeq\sigma_M/H$ and
$\alpha_1\simeq\sigma_M/H^2$.
Using the scaling result for $\alpha_e$ in this regime, we find that
the effective permittivity is given by
\begin{equation}
\label{epsehighH}
\varepsilon_e\simeq\frac{q_c}{3}\left(\frac{\varepsilon_I}{|\Delta
p|}-i\frac{Aq_c^2}{12\pi}\frac{\omega\varepsilon_I^2}{\sigma_M|\Delta
p|^3}
\right).
\end{equation}
This scaling behavior is the same as what was found above in the
weak field regime. The first term should indeed be independent of $H$,
since it corresponds to the universal behavior of the d.c.\
permittivity near the percolation threshold. However, the fact that the
imaginary parts of (\ref{epsehighH}) and (\ref{epselowH}) are the same
to the order shown here is accidental, and is probably
due to the nature of the approximations used (a similar accident
also occurs in the SEMA results). Indeed, we can see from
(\ref{epstilda-st}) that these parts depend on the current distribution
in the metallic inclusions, hence they should be different in the weak
and strong field regimes. The important physical conclusion is that,
in both regimes, $\varepsilon_e$ is independent of $H$.

In the strong field regime, we obtain from (\ref{betahighH}) the
surprising result
\be
\label{epstildehighH}
\tilde{\varepsilon}_e
\propto -\frac{\omega\varepsilon_I^2}{4\pi\sigma_M (\Delta p)^2}H,
\ee
which is again the same scaling behavior as in the weak field regime.
In both regimes, the Faraday coefficient thus reads
\be
\label{faracoeffreplica}
{\cal F}\propto -\frac{\omega\varepsilon_I^{3/2}H}{4\pi\sigma_M|\Delta
p|^{3/2}}.
\ee
This means that the Faraday rotation (the angle is
proportional to
${\cal F}$) is proportional to the applied
magnetic field, even for strong fields, i.e., when
$H\gg 1$. It is thus clear that
we can obtain a large value of the rotation angle using such a
composite.

We now consider the results obtained by means of the
high contrast or
low frequency expansion. The scaling behavior of the averages which
appear in
(\ref{hat_epsilon_e}) can be deduced by comparing
(\ref{rho_e_I_M}) to scaling theories previously developed for
d.c. magneto-transport coefficients
\cite{BergmanStroud85,SarBergStrelPRB93}, and from the
property that
\begin{equation}
{\bf J}^{(0)}=\nabla\times{\bf A}^{(0)}\propto
\left\{\begin{array}{cll}
1 & {\bf r}\in\mbox{ percolating cluster;} & p>p_c\\
0 & {\bf r}\in\mbox{ elsewhere;} & p>p_c\\
1 & {\bf r}\in\mbox{ anywhere;} & p<p_c.
\end{array}\right.
\end{equation}
These considerations lead to
\begin{eqnarray}
\langle\theta_I(\nabla\times{\bf A}^{(0)})\rangle&\propto&
\left\{\begin{array}{lll}
\Delta p^s; & p<p_c, & \mbox{any $H$} \\
0; & p>p_c, & \mbox{any $H$} \end{array}\right.\\
\langle\theta_M(\nabla\times{\bf A}^{(0\alpha)})^2\rangle&\propto&
\left\{\begin{array}{lll}
\Delta p^{-t}; & \mbox{any $p$} & H\ll 1\\
\Delta p^{-t_H}F\left(\xi_p\over\xi_H\right); & \mbox{any $p$} & H\gg 1
\end{array}\right.\\
\langle\theta_M[(\nabla\times{\bf A}^{(0x)})\times
(\nabla\times{\bf A}^{(0y)})]_z\rangle&\propto&
\left\{\begin{array}{lll}
\Delta p^{-g}; & \mbox{any $p$} & H\ll 1\\
\Delta p^{-g_H}G\left(\xi_p\over\xi_H\right); & \mbox{any $p$} & H\gg 1
\end{array}\right..
\end{eqnarray}
Here $\xi_H=H^{\nu_H}$ is the magnetic field dependent correlation
length, $\xi_p\propto\Delta p^{-\nu}$ is the percolation
correlation length, and $F(z)$, $G(z)$ are scaling functions,
which tend to nonzero constants when $z\ll 1$, and to asymptotic
forms that return the $\Delta p$ exponents to their $H\ll 1$
values when $z\gg 1$ (see Ref. \onlinecite{SarBergStrelPRB93}):
\begin{eqnarray}
F(z)&\propto&\left\{\begin{array}{ll}
{\rm const,} & z\ll 1\\
z^{t-t_H\over\nu}, & z\gg 1\end{array}\right.\\
G(z)&\propto&\left\{\begin{array}{ll}
{\rm const,} & z\ll 1\\
z^{g-g_H\over\nu}, & z\gg 1.\end{array}\right.
\end{eqnarray}
The values of the critical exponents which appear in the
above scaling expressions, as determined by simulations
of three dimensional percolating network models, are
\cite{AharonyStauffer,Duering,SarBergStrelPRB93}
\begin{equation}
\begin{array}{llll}
\nu\cong 0.88, & t\cong 2.0, & s\cong 0.7, & g\cong 0.38,\\
\nu_H\cong 0.5, & t_H\cong 6.0, &  & g_H\cong 5.0.
\end{array}
\end{equation}

Finally, we get the following results for
the Faraday coefficient of a percolating mixture below $p_c$
\begin{equation}
-\frac{\omega\varepsilon_I^{3/2}H}{4\pi\sigma_M}\,\frac{\langle\theta_M
[(\nabla\times{\bf A}^{(0x)})\times(\nabla\times{\bf A}^{(0y)})]_z\rangle}
{\langle\theta_I(\nabla\times{\bf A}^{(0x)})_x\rangle^{3/2}}\propto
\left\{\begin{array}{ll}
-\frac{\omega\varepsilon_I^{3/2}H}{4\pi\sigma_M}\Delta p^{-g-3s/2},
& H\ll 1\\&\\
-\frac{\omega\varepsilon_I^{3/2}H^{1+(g_H-g)\nu_H/\nu}}
{4\pi\sigma_M}\Delta p^{-g-3s/2},& H\gg 1\;
\;{\rm but}\;\xi_H\ll \xi_p\\
-\frac{\omega\varepsilon_I^{3/2}H}{4\pi\sigma_M}\Delta p^{-g_H-3s/2},
& \xi_H\gg\xi_p.\end{array}\right.
\label{Faraday_coeff}
\end{equation}

These results are consistent with (\ref{faracoeffreplica}), which was
obtained using the replica method, where we expect to find the SEMA values
$s=1$, $g=g_H=0$. We can also derive a number of physical consequences
which follow from both approaches [from (\ref{epstilda-st}) in the
high contrast expansion, and from (\ref{epselowH}) and (\ref{epsehighH})
in the replica approach]:
\begin{enumerate}
\item The diagonal part of $\hat\varepsilon_e$ has an imaginary part
that is proportional to $\omega\varepsilon_I^2/\sigma_M$ and is
independent of $H$ (up to $O(\rho_M)$ in the high contrast
expansion). This means that there will be some dissipation.
\item In the same order, $\hat\varepsilon_e$ has an
antisymmetric part which is imaginary and proportional to
$\omega\varepsilon_I^2 H/\sigma_M$. These results are valid both for
$H<1$ (weak field) and for $H>1$ (strong field), as long as both
$\omega\varepsilon_I\ll\sigma_M$ and $\omega\varepsilon_I H\ll\sigma_M$.
\end{enumerate}

We note that the scaling predictions for magneto-transport in a
percolating system have been tested experimentally only for
weak fields, and only in systems that were above the percolation
threshold $p_c$ \cite{Dai}.
Measurements of the induced Faraday effect in a metal-dielectric
mixture below $p_c$ could therefore provide an important
test of those predictions. Another
prediction which follows from (\ref{hat_epsilon_e})
is that the induced
Faraday effect in a non-conducting metal-dielectric composite,
which is not necessarily near any percolation threshold, is
linear in $H\equiv\omega_c\tau$ even when $H\gg 1$,
in agreement with the replica approach near $p_c$.


\section{Conclusion}
\label{conclusion}

In this paper, we studied the Faraday effect (in the
quasi-static limit) in a composite consisting of a dielectric
matrix (with negligible Faraday effect) with metallic
inclusions (which have only Hall effect and no intrinsic
magneto-resistance). We presented two different approaches
leading essentially to the same conclusions. The first
approach relies on the replica method and allowed us to
derive in a non-perturbative way equations for the
effective permittivity tensor. The second approach is the
result of an expansion in powers of $\rho_M/\rho_I$
[$\rho_M$ ($\rho_I$) is the impedance of the metallic (dielectric)
component], combined with scaling ansatzes near $p_c$.

First of all, both approaches are consistent with each other, the only
difference is that the scaling exponents predicted by the replica approach
have their SEMA values.
Second, the main result is the following: the scaling of the Faraday
coefficient is the same for the weak field and the strong field
regimes (as long
as both
$\omega\varepsilon_I\ll\sigma_M$ and $\omega\varepsilon_I H\ll\sigma_M$).
In particular, we found that
the Faraday angle (proportional to ${\cal F}$)
is proportional to the magnetic field {\bf B} even for
strong field, as long as $p<p_c$ and $\frac{\varepsilon_I\omega}
{\sigma_M\Delta p^2}\ll 1$. We can thus predict that it should be
in principle possible to obtain large values of the rotation angle in
such a system. For instance, for $\Delta p$ of order $0.1$ (which
is realistic in experiments), $\varepsilon_I$ of order unity and
$\sigma_M/\omega$ of order $100$ for semiconductors, and with
$\omega$ in the microwave region (the ratio $\sigma_M/\omega$
should not be too large since $\tilde{\varepsilon}_e$
is proportional to its
inverse), one obtains for the Faraday coefficient
\begin{equation}
\frac{\tilde{\varepsilon}_e}{\sqrt{\varepsilon}_e}\simeq 10^{-1} H,
\end{equation}
which can be made of order unity using currently available
magnetic fields and high mobility doped semiconductors. One should
recall that in homogeneous dielectrics, the Faraday coefficient
is usually much less than 1: Typical values for a $1$\,T magnetic
field, and for a wavelength in the visible spectrum ($\lambda\simeq
0.6 \mu$m), are of order
$10^{-6}$ for dielectrics like quartz, and of order $10^{-2}$ for
thin ferromagnetic metallic iron films. Measurements of the
Faraday effect
below $p_c$ in a percolating
metal-dielectric composite could provide an important test of the scaling
predictions in both the strong field and the weak field regimes.
Such experiments would have to involve either propagation or reflection
of microwaves by a metal-dielectric composite with metallic
inclusions that are smaller than the relevant skin depth.
Both approaches also predict that the transverse diagonal elements of
$\hat\varepsilon_e$ have an imaginary part that is proportional to
$\omega\varepsilon_I^2/\sigma_M$, and are independent of $H$
[up to terms of order $O(\rho_M)$ in the
high contrast expansion]. This means that there will be
some dissipation.


\acknowledgements

One of us (MB) acknowledges hospitality from Tel-Aviv University where
this work was finished. This research was supported in part by grants
from the
US-Israel Binational Science Foundation, the Israel Science Foundation,
and the Tel Aviv University Research Authority.

\appendix
\section{}

We want here to solve the Maxwell equation satisfied by the electric
field {\bf E} (where $k_0=\omega/c$)
\begin{equation}
[\nabla\times\nabla\times+\hat{\varepsilon}({\bf r})k^{2}_{0}]
{\bf E}({\bf r})=0.
\end{equation}
The tensor $\hat{\varepsilon}$ is a random variable equal to
$\hat{\varepsilon}_1$ with probability $p$ and to $\hat{\varepsilon}_2$
with probability $q=1-p$ [$\hat{\varepsilon}_1$ and $\hat{\varepsilon}_2$ are tensors of the form given in (\ref{matrice})]. 
In order to obtain an integral equation for the electric field,
we first write $\hat{\varepsilon}({\bf r})=\varepsilon_{0}\hat{I}+
\delta\hat{\varepsilon}({\bf r})$ (where $\varepsilon_0$ is an arbitrary
constant which will disappear at the end of the calculation). One
can then easily show that ${\bf E}({\bf r})$ is also the solution of
the following equation
\begin{equation}
\label{inteqforE}
{\bf E}({\bf r})={\bf E}_{0}({\bf r})+\int{\rm d}{\bf r}'\hat{G}
({\bf r}-{\bf r}')
\delta\hat{\varepsilon}({\bf r}'){\bf E}({\bf r}'),
\end{equation}
where $\hat{G}$ is the dipolar tensor for the uniform medium
of permittivity $\varepsilon_0$ and where the quantity ${\bf E}_0$
depends only on the boundary conditions and is assumed to be
uniform.

We will use the Fourier transform of  $G_{\alpha\beta}({\bf r})$ given by
\begin{equation}
\label{fourierG}
G_{\alpha\beta}(k)=-{k_{\alpha}k_{\beta}\over{\varepsilon_{0}k^{2}}}+
{k_{0}^{2}\over{k^{2}-\varepsilon_{0}k_{0}^{2}}}[\delta_{\alpha\beta}-
{k_{\alpha}k_{\beta}\over{k^{2}}}].
\end{equation}
We work in the quasi-static limit which
means that we take the limit $k_0$ going to zero. In this case, the
dipolar tensor is given by
\begin{equation}
\label{fourierG2}
G_{\alpha\beta}(k)=-{k_{\alpha}k_{\beta}\over{\varepsilon_{0}k^{2}}}
\;\;{\rm for}\;\;k\neq 0.
\end{equation}
At $k=0$, the value of this tensor is 
$G_{0}=-\delta_{\alpha,\beta}/(d\varepsilon_0)$, where $d$ is the
space dimension, here equal to three.
For values of $k_0$ that are too large,
we cannot define an effective permittivity
tensor and we have to introduce the notion of spatial dispersion (for
a review, see e.g., Ref.\ \onlinecite{klyu}
and for a study using the replica method
see Ref.\ \onlinecite{boz}).

Averaging equation (\ref{inteqforE}), after inverting it, and averaging
it before inverting it leads to the following exact relation
\begin{equation}
\label{Mminus1}
<M^{-1}_{\alpha\beta}(k=0)>=(1-G_{0}\delta\hat{\varepsilon}_{e})^{-1}_
{\alpha\beta},
\end{equation}
where $\hat{M}^{-1}$ is the inverse of the random operator
\begin{equation}
M_{\alpha\beta}({\bf r},{\bf r}')=\delta_{\alpha\beta}\delta({\bf
r}-{\bf r}')-
(\hat{G}({\bf r}-{\bf r}')\delta\hat{\varepsilon}({\bf
r}'))_{\alpha\beta}.
\end{equation}

Calculation of the effective permittivity tensor
is thus reduced to a calculation of the average over the disorder
of the inverse of a random operator, namely $\hat{M}$. In order
to do this, we will use the replica method which allows us to
express the elements of $\hat{M}^{-1}$ (after using a Gaussian inversion
formula) in terms of the functional integral
\be
\label{replica1}
M^{-1}_{\alpha\beta}({\bf r},{\bf r}')=\int {\cal D}(\bar{\psi} ,\psi)
\bar{\psi}_{\alpha}^{a}({\bf r})
\psi_{\beta}^{a}({\bf r}')
e^{\int{\rm d}{\bf r}{\rm d}{\bf r}'
\sum_{\alpha\beta,a}\bar{\psi}_{\alpha}^{a}({\bf r})M_{\alpha\beta}
({\bf r},{\bf r}')
\psi_{\beta}^{a}({\bf r}')},
\ee
where $\psi^{a}_{\alpha}$ (with $a=1,...,n$)
(and its conjugate $\bar{\psi}$) are replicated Grassman fields
satisfying the usual anticommutation relations
\be
\{\psi({\bf r}),\psi({\bf r}')\}=\{\psi({\bf r}),\bar\psi({\bf r}')\}
=\{\bar\psi({\bf r}),\bar\psi({\bf r}')\}=0.
\ee

The limit $n=0$ is implicitely taken in (\ref{replica1})
and, as usual in the replica method, we first
consider $n$ as an integer and then take the limit $n$ going to zero
at the end of the calculation (without adressing the problem
of analytic continuation).

It is now easy to average $\hat{M}^{-1}$ over the disorder, and we
obtain
\begin{equation}
\label{replica2}
\langle M^{-1}\rangle_{\alpha\beta}({\bf r-r'})=\int {\cal D}(\bar\psi
,\psi)
\bar\psi_{\alpha}^{a}({\bf r})
\psi_{\beta}^{a}({\bf r}')
e^{{\cal H}_e},
\end{equation}
where the effective Hamiltonian is given by
\begin{eqnarray}
{\cal H}_{e}=\int{\rm d}{\bf r}{\rm d}{\bf r}'\sum_{\alpha,\beta,a}
\bar{\psi}_{\alpha}^{a}({\bf r})
\{\delta_{\alpha\beta}\delta({\bf r}-{\bf r}')-
(\hat{G}({\bf r}-{\bf r}')\delta\hat{\varepsilon}_1)_{\alpha\beta}\}
\psi_{\beta}^{a}({\bf r}')
\nonumber\\
+\int{\rm d}{\bf r}_0\ln
\left[
1+\eta
e^{
\int{\rm d}{\bf r}'\sum_{\alpha,\beta,\mu,a}
\bar{\psi}_{\alpha}^{a}({\bf r}_0)G_{\alpha\mu}({\bf r}_0-{\bf
r}')\Delta_{\mu\beta}
\psi_{\beta}^{a}({\bf r}')}
\right].
\end{eqnarray}

The matrix $\hat{\Delta}$ is equal to
$\hat{\varepsilon}_2-\hat{\varepsilon}_1$ and
$\eta=(1-p)/p$ (note that $\eta$ is the inverse of $\lambda=p/q$ which appears in the main text). As usual in the replica method, the average over
disorder introduces coupling between different replicas (if there
is no coupling, then the averaging is trivial) and in order
to study this complicated effective Hamiltonian, we will use
a variational principle \cite{variat,bo}.
This principle consists of finding the best Gaussian
approximation ${\cal H}_0$ to the effective Hamiltonian ${\cal H}_e$. Denoting by $\hat{K}^{-1}$
the kernel of ${\cal H}_0$ (the variational approximation thus reads
$\langle\hat{M}^{-1}\rangle\simeq\hat{K}$), we
have to minimize with respect to $\hat{K}$ the following variational
free energy $\Phi$
\begin{equation}
\Phi(\hat{K})={\cal F}_0+\langle{\cal H}_e-{\cal H}_0\rangle_0,
\end{equation}
where ${\cal F}_0$ is the free energy associated with ${\cal H}_0$ and
where $\langle\cdot\rangle_0$ denotes an average using ${\cal H}_0$. We
thus
obtain the following equation
\be
\label{variaeq}
(\hat K^{-1})_{\alpha\beta}=
\delta_{\alpha\beta}-(\hat{G}(k)\delta\hat{\varepsilon}_e)_
{\alpha\beta},
\ee
with
\begin{eqnarray}
\label{tensoreff}
\hat{\varepsilon}_e=\hat{\varepsilon}_1+\sum_{m\ge 1}
(-)^{m+1}\eta^{m}
\frac
{\hat{\Delta}}
{1-m\hat{\Delta}\int\frac{{\rm d}^{d}k}{(2\pi)^{d}}\hat{K}\hat{G}}.
\end{eqnarray}

The tensor $\hat{K}$ can be inverted and we obtain
\begin{equation}
\hat{K}=\frac{1}{1+\frac{{\bf k}\cdot{\bf q}}{D}}
\left[1+\frac{{\bf k}\cdot{\bf q}}{D}-\frac{{\bf k}\otimes{\bf
q}}{D}\right],
\end{equation}
where $D=\varepsilon_{0}k^{2}$, ${\bf
q}=\delta\hat{\varepsilon}^{t}_{e}{\bf k}$
($\delta\hat{\varepsilon}^{t}_{e}$ is the transpose of
$\delta\hat{\varepsilon}_{e}$)
and where $\otimes$ denotes the usual dyadic product. We can then compute
$\hat{K}\hat{G}$ and we find that it is the dipolar tensor for the
effective
medium
\begin{equation}
(\hat{K}\hat{G})_{\alpha\beta}=-\frac{k_{\alpha}k_{\beta}}
{{\bf k}\cdot(\varepsilon^{t}_{e}{\bf k})}
\nonumber\\
=-\frac{k_{\alpha}k_{\beta}}
{\sum_{\alpha=1}^{3}(\varepsilon_{e})_{\alpha\alpha}k_{\alpha}^{2}}.
\end{equation}

It is then easy to integrate $\hat K\hat G$ and we obtain
\begin{equation}
\int\frac{{\rm d}{\bf k}}{(2\pi)^{3}}(\hat{K}\hat{G})_{\alpha\beta}=
-{1\over 3}\delta_{\alpha\beta}\mu_{\alpha}=-\frac{1}{3}\mu_{\alpha\beta},
\end{equation}
where $\hat{\mu}$ is a diagonal matrix with diagonal elements
 $\mu_1=\mu_2=1/\varepsilon_{e}$ and $\mu_3=1/\varepsilon_{ze}$.
The self-consistent equation (\ref{tensoreff}) can thus be recast as
the following matrix equation
\begin{equation}
\hat{\varepsilon}_e=\int_{0}^{\infty}{\rm d}ue^{-u}
\frac{\langle\hat{\varepsilon}e^{-u\hat{\varepsilon}\hat{\mu}/3}\rangle}
{\langle e^{-u\hat{\varepsilon}\hat{\mu}/3}\rangle},
\end{equation}
where the brackets still denote an average over the disorder. It is then
easy to show that
\begin{equation}
\left(\begin{array}{ccc}
\varepsilon_e & i\tilde{\varepsilon}_e \\
-i\tilde{\varepsilon}_e & \varepsilon_e \\
\end{array}\right)=
\int_{0}^{\infty}{\rm d}ue^{-u}
\frac{\langle\hat{\varepsilon}_{2}e^{-u\hat{\varepsilon}_{2}/3\varepsilon_e}\rangle}
{\langle e^{-u\hat{\varepsilon}_{2}/3\varepsilon_e}\rangle},
\end{equation}
where $\hat{\varepsilon}_2$ denotes the restriction of the tensor $\hat{
\varepsilon}$ to the $(x,y)$ subspace
\begin{equation}
\hat{\varepsilon}_{2}=
\left(\begin{array}{ccc}
\varepsilon & i\tilde{\varepsilon} \\
-i\tilde{\varepsilon} & \varepsilon \\
\end{array}\right).
\end{equation}

The equation along the $z-$axis is decoupled from the preceeding one and
is  Eq.\ (\ref{epsz}) of the main text. Using the following relation
\be
e^{\alpha\hat{\varepsilon}_{2}}=
e^{\alpha\varepsilon}
\left(\begin{array}{ccc}
\cosh\alpha\tilde{\varepsilon} & -i\sinh\alpha\tilde{\varepsilon} \\
i\sinh\alpha\tilde{\varepsilon} & \cosh\alpha\tilde{\varepsilon} \\
\end{array}\right),
\ee
we obtain after simple manipulations equations (\ref{epseff1}) and
(\ref{epstilde}) of the main text.
It should be noted that all the calculation presented here can be used
without any changes for either real or complex values of $\varepsilon$ and
$\tilde{\varepsilon}$. This justifies the analytical continuation
of the formulas obtained in the framework of the Hall effect in
order to describe the Faraday effect.


\section{}

The equations for the Green tensor $\hat G^{(\rho)}({\bf r},{\bf r'})$
can be solved in almost closed form if the system occupies all
space, and if the resistivity tensor $\hat\rho$ is constant everywhere
and its symmetric part is a scalar tensor, i.e., if
\begin{equation}
\hat\rho({\bf r})\cdot{\bf v}\equiv\hat\rho_0\cdot{\bf v}=
\rho_0{\bf v} + {\bf b}\times{\bf v}\label{rho_cross_product}
\end{equation}
for any vector {\bf v}. In that case, the Green tensor depends only on
${\bf r}-{\bf r'}$, and we can define its Fourier transform by
\begin{equation}
\tilde G^{(\rho_0)}_{\alpha\beta}({\bf q})\equiv\int {\rm d}({\bf
r}-{\bf r'})
\,G^{(\rho_0)}_{\alpha\beta}({\bf r}-{\bf r'})
e^{-i{\bf q}\cdot({\bf r}-{\bf r'})}.\label{G_Fourier}
\end{equation}
Using (\ref{G_eq}), it is easily found that this Fourier transform
satisfies the following linear algebraic equations
\begin{equation}
(\rho_0q^2-k^2)\tilde G^{(\rho_0)}_{\alpha\beta}
-\rho_0q_\alpha({\bf q}\cdot\tilde G^{(\rho_0)}_{\cdot\beta})
+({\bf b}\cdot{\bf q})({\bf q}\times\tilde
G^{(\rho_0)}_{\cdot\beta})_\alpha
=\delta_{\alpha\beta},\label{G_q_eq}
\end{equation}
which can be solved to yield ($\varepsilon_{\alpha\beta\gamma}$ is
the basic antisymmetric tensor)
\begin{equation}
\tilde G^{(\rho_0)}_{\alpha\beta}({\bf q})=
{(\rho_0q^2-k^2)\delta_{\alpha\beta}-[\rho_0(\rho_0q^2-k^2)+
({\bf b}\cdot{\bf q})^2]\,q_\alpha q_\beta/k^2 +
({\bf b}\cdot{\bf q})\varepsilon_{\alpha\beta\gamma}q_\gamma\over
(\rho_0q^2-k^2)^2 + q^2({\bf b}\cdot{\bf q})^2}.\label{G_q_solution}
\end{equation}
Clearly, $\tilde G^{(\rho_0)}_{\alpha\beta}({\bf q})$ diverges
in the limit $k\rightarrow 0$. However, if one calculates the
Fourier transform of $(\nabla\times G^{(\rho_0)}_{\cdot\beta})_\alpha$,
namely
\begin{equation}
({\bf q}\times\tilde G^{(\rho_0)}_{\cdot\beta})_\alpha=
{({\bf b}\cdot{\bf q})(q^2\,\delta_{\alpha\beta}-q_\alpha q_\beta)
-(\rho_0q^2-k^2)\varepsilon_{\alpha\beta\gamma}q_\gamma\over
(\rho_0q^2-k^2)^2 + q^2({\bf b}\cdot{\bf q})^2},\label{q_cross_G}
\end{equation}
then the limit $k\rightarrow 0$ can be taken without any problems.
That is why we had to include the term $k^2\,G^{(\rho)}_{\alpha\beta}$
in the equation for the Green tensor [see (\ref{G_eq})], deferring
the limit $k\rightarrow 0$ until after the calculation of
$(\nabla\times G^{(\rho)}_{\cdot\beta})_\alpha$.

In order to investigate the symmetry properties of
$\hat G^{(\rho)}({\bf r},{\bf r'})$, we use integration by parts
or Green's theorem to get, for any vector fields
${\bf A}({\bf r})$, ${\bf B}({\bf r})$, and second rank tensor
field $\hat\rho({\bf r})$,
\begin{eqnarray}
\lefteqn{\int_V {\rm d}{\bf r}\left\{{\bf
A}\cdot\nabla\times[\hat\rho\cdot
(\nabla\times{\bf B})]-{\bf B}\cdot\nabla\times[\hat\rho\cdot
(\nabla\times{\bf A})]\right\}}\nonumber\\
&=&-\oint_{\partial V}\left[(d{\bf S}\times{\bf A})\cdot
\hat{\rho}\cdot(\nabla\times{\bf B})-
(d{\bf S}\times{\bf B})\cdot\hat\rho\cdot(\nabla\times{\bf A})\right]
\nonumber\\&&
+\int_V {\rm d}{\bf r}\left[(\nabla\times{\bf
A})\cdot\hat\rho\cdot(\nabla\times{\bf B})
-(\nabla\times{\bf B})\cdot\hat\rho\cdot(\nabla\times{\bf A})\right].
\label{Green_theorem}
\end{eqnarray}
If $\hat\rho$ is {\em symmetric}, then the integrand in the last
volume integral vanishes everywhere. Substituting
\begin{equation}
A_\omega({\bf r})\equiv G^{(\rho)}_{\omega\alpha}({\bf r},{\bf r}_1),
\;\;\;B_\omega({\bf r})\equiv G^{(\rho)}_{\omega\beta}({\bf r},{\bf r}_2)
\label{substitution}
\end{equation}
in this result, and assuming that $\hat\rho$ is a symmetric tensor,
we thus get, using (\ref{G_eq}) and (\ref{G_bound_cond}),
\begin{equation}
G^{(\rho)}_{\alpha\beta}({\bf r}_1,{\bf r}_2)=
G^{(\rho)}_{\beta\alpha}({\bf r}_2,{\bf r}_1)\;\;\;{\rm if}\;\;
\hat\rho^t=\hat\rho.\label{reciprocity}
\end{equation}
Note that if $\hat\rho$ is non-symmetric, then in general
$G^{(\rho)}_{\alpha\beta}({\bf r},{\bf r'})$ is not a symmetric
kernel.



\end{document}